# Open Access Scientometrics and the UK Research Assessment Exercise

Stevan Harnad
Institut des sciences cognitives
Université du Québec à Montréal
Montréal, Québec Canada H3C 3P8
http://www.crsc.uqam.ca/
and
Department of Electronics & Computer Science
University of Southampton
Highfield, Southampton, UK SO17 1BJ
http://www.ecs.soton.ac.uk/~harnad/

**ABSTRACT:** Scientometric predictors of research performance need to be validated by showing that they have a high correlation with the external criterion they are trying to predict. The UK Research Assessment Exercise (RAE) – together with the growing movement toward making the full-texts of research articles freely available on the web -- offer a unique opportunity to test and validate a wealth of old and new scientometric predictors, through multiple regression analysis: Publications, journal impact factors, citations, co-citations, citation chronometrics (age, growth, latency to peak, decay rate), hub/authority scores, h-index, prior funding, student counts, co-authorship scores, endogamy/exogamy, textual proximity, download/co-downloads and their chronometrics, etc. can all be tested and validated jointly, discipline by discipline, against their RAE panel rankings in the forthcoming parallel panel-based and metric RAE in 2008. The weights of each predictor can be calibrated to maximize the joint correlation with the rankings. Open Access Scientometrics will provide powerful new means of navigating, evaluating, predicting and analyzing the growing Open Access database, as well as powerful incentives for making it grow faster.

Scientometrics probably began as Learned Journals took over from peer-to-peer scholarly letter-writing in the 17th century (Guédon 2002), but it came into its own in our own 'publish-or-perish' era. With Garfield's (1955) contibutions to citation counting and indexing in the 1950's, the academic bean-counting of publications for performance evaluation and funding came to be supplemented by citation counting. It was no longer enough just to publish in bulk: it had to be demonstrable that your publications were also heavily used, hence useful and important. A direct indicator of usage was the fact that your research was cited by subsequent research (Moed 2005).

Of course 'heavy' varies with the field. There are industrial-scale research areas and sparse esoteric ones, having only a few peers worldwide (rather as in the letter-writing era). So it was always true (if not always taken into account) that citation counts would have to be used with caution, always comparing like with like rather than turnips with truffles. It makes no sense to point out that a paper or author in cancer research has more citations than a paper or author in Finno-Ugric philology. Nor that a full professor in the fullness of his years has more citations than a fresh post-doc. Journals too are hard to compare, unless their subject matter is closely equated. And although for a while it became the most popular bean-counting method, using the 'journal impact factor' (the average number of citations per article) to weigh publications is like using the average graduating marks of their secondary schools to weigh incoming university applicants. One wants the applicant's own exact marks, and then one wants to know what those marks mean.

**Psychometrics and Test Validation.** In weighing the meaning of metrics, scientometrics can perhaps take a lesson from psychometrics (Kline 2000). In the field of aptitude testing, the tests are constructed and validated against external criteria. One first starts by inventing test items, picking items that agree with one another in polarity: A set of items is given to a large number of testees, and their average scores on half of the test items are compared with their average scores on the other half. If the correlation is not very high, then the test does not even agree with itself, let alone predict something external to itself. Split-half correlations are called a test's reliability. Test/re-test correlations are a further measure of reliability.

Now suppose we have a reliable test, with split-halves and test/re-test being highly correlated: Is the test valid? Does it measure anything other than itself ? After all, in aptitude testing, we cannot afford to define aptitude 'operationally' -- as simply amounting to the score on the reliable test that I happen to have constructed and baptized a test of aptitude. Aptitude tests do not have face validity. We have to show that their score is predictive, in that it correlates with something else that we already agree to call aptitude.

In this respect, aptitude testing is like weather-forecasting. It is not enough to show that measures of barometric pressure are reliable, and tend to agree with themselves when measured repeatedly or in different ways. It has to be shown that they predict rain. If there is a high correlation between barometric pressure and probability of precipitation, then barometric pressure is validated as a predictor of impending rain. In the same way, we can correlate our aptitude test scores with some external measure of aptitude (age-related school performance norms perhaps, or postgraduation job performance, or human judgments of individuals' relative intelligence). This external variable against which psychometricians validate their test scores is called the 'criterion', and the measure of the validity of a test is essentially the size of its correlation with its criterion (or criteria).

Now note the difference between psychometics and meteorology: Barometric pressure either does or does not correlate with the likelihood of rain. If it doesn't, then we need to look for another predictor. In the case of aptitude tests, if we have an unreliable test, one that does not agree with itself, we can keep constructing new items and discarding old ones until we have made the test internally reliable. Then, if a reliable test has a low correlation with its external criterion, we can try discarding old items and constructing new ones in order to try to raise the test's correlation with its criterion to the level of predictive validity we need for normative use. In meteorology, if the correlation between pressure and rain is not high enough, we can't change the measure of pressure, but we can add more predictors, for example, current humidity, or the speed and direction of an approaching cold front. The measures can then be combined in a [multiple regression](multiple regression) equation, in which, say, 3 predictors, each with its own regression (or 'beta') weight, are used jointly to predict the criterion.

The same can be done with psychometric test validation. If the validity of a single test is too low, we can add more tests : tests of verbal ability, tests of numerical ability, tests of spatial ability – to see whether, jointly, they do a better job of predicting our criterion.

**Multivariate Metrics.** Now contrast this with the case of scientometrics, and citation counting in particular. Although I don't know whether it has been tested formally, let us assume that citation counts are indeed reliable, in that if you randomly split an author's works in half, there will be a good split-half correlation, and that if you take successive citation counts in two different time windows, they will be sufficiently correlated with one another too.

Now we proceed to validation : Correlations between citations and various criteria (e.g., performance indicators, such as funding, productivity, prizes) have been measured, much the way correlations between psychometric scores and their criteria have been measured. And those scientometric correlations have turned out to be whatever they have been : sometimes higher sometimes lower, but never, I would suggest, stellar – never approaching the kind of predictivity there is in meteorology or aptitude psychometrics or medical biometrics. And the reason validity is not higher is that we cannot really discard old test items and construct new ones to improve the correlation of citations with their criterion, as we do in psychometrics. Citation counts are citation counts. So we have thus far been rather passive about the validation of our scientific and scholarly performance metrics, taking pot-luck rather than systematically trying to increase their validity, as in psychometrics.

For a long time we simply stuck superstitiously with the journal citation impact factor. As noted, in evaluating individual papers or authors the average citation count of the journal in which their paper was published is a rather blunt instrument. Performance evaluation committees have since begun to look also at exact citation counts for papers or authors. There have been some gestures toward using co-citations too (after all, it should make a difference if a paper or author is co-cited with a graduate student versus a Nobel Laureate). The Institute for Scientific Information developed an 'immediacy index' for journals ; in principle, similar time-based measures could index an individual paper's or author's citation growth, latency to peak, and decay rate. Let's call those 'citation chronometrics'. Then there were 'hub' and 'authority' scores, measuring citation fan-out and fan-in (Kleinberg 1999). The authority score is a measure similar to Google's PageRank, that recursively weights incoming citations by their own respective incoming citation weights (Page et al. 1999). Lately we have also had Hirsch's (2005) h-index. Online download counts have likewise entered the arena. Many other metrics are possible too: co-authorship, endogamy/exogamy (how narrow or wide and even cross-disciplinary is a paper's or author' citation fan-in, fan-out, and wider interconnectivity?), textual proximity (degree of textual overlap, and latent semantic distance metrics ; Landauer et al 1998), download chronometrics, perhaps eventually even co-download measures. But for some reason, we have so far tended to use these metrics one at a time, as if both predictor and criterion were univariate, rather than trying to combine them into multivariate measures that might have higher predictive power.

The first thing psychometricians would do with a 'battery' of univariate metrics would be to systematically validate them against external criteria that already have some face validity for us : There are other classic performance measures, such as funding, doctoral student counts, and prizes -- but, frankly, using those would be circular, as they have not been externally validated either. What psychometricians sometimes do first with their batteries of diverse metrics is to look at their intercorrelational structure through principal-component and factor analyses. What has emerged, somewhat surprisingly, from factor-analyzing many diverse aptitude tests jointly is a single primary factor – the General Intelligence or 'G' factor – that is common to all of them. There are those who think G is some sort of artifact of the test construction and correlational methods used, but most psychometricians think G is actually an empirical finding : that there is a single basic human intellectual ability underlying all the other special abilities (Kline 2000).

Aptitude tests differ in the size of their G 'loadings', but all of them have some positive G loading. No one test, however, is a direct measure of G alone. It requires a battery of tests to get a picture of G (and so far individuals are not characterised as having a given 'G Score', but as having a composite of scores on a variety of different aptitude tests). Intelligence is multidimensional even if there is a shared underlying factor. Hence aptitude tests have to be multidimensional. But aptitude testing has the advantage over scientometrics that most of the various specific aptitude tests (verbal reasoning, mathematical reasoning, spatial visualization, short-term memory, reaction time, musical ability, motor coordination, etc., some of them psychometric, some of them biometric) have been validated against their respective criteria. This is not true in scientometrics: Citation and other metrics have been used, and their pairwise correlations with criteria have been calculated and reported, but nothing like the systematic validation process that goes into the construction and use of aptitude tests – calibrating and optimizing on the basis of data from generation after generation of students -- has been done with scientometric measures. Nothing like standardized 'norms' or benchmarks has as yet emerged from scientometrics.

It has to be said that this is partly because the database containing the most important of the scientometric indicators – citations -- has been in the proprietary hands of one sole database provider for decades, with parts of it temporarily leased (at no small cost) to those who wished to do some data-mining and analyses. This is about to change, with the onset of the 'Open Access' era:

**Open Access.** Until now, the reference metadata and cited references of the top 25% of the c. 24,000 peer-reviewed journals published worldwide, across disciplines and languages, have been systematically fed (by the journal publishers) to the Institutite for Scientific Information (ISI), to be extracted and stored. But soon this is will change. It has been discovered (belatedly) that the Web makes it possible to make the full-text (not just the reference metadata and cited reference) of every single one of the 2.5 million articles published annually in those 24,000 journals (not just the top 25%) freely accessible online to all users (not just those that can afford paid access to the journals and the ISI dtabase).

Open Access (OA) means free online access to all peer-reviewed journal articles. In the paper era, because of the cost of generating and disseminating print-on-paper, OA was unthinkable. Paper access could not be provided free for all, because subscription income was needed to cover the costs of peer review and publishing. Moreover, the paper medium did not make it possible to (literally) put the entire peer-reviewed journal corpus at the fingertips of all users, everywhere, at all times (in the way the Web already does today – but with other forms of content, such as E-bay product blurbs, blogs, and pornography).

Awareness of the Web's potential for providing OA to the research corpus is arriving belatedly for [a number of reasons](), but the chief two reasons are that for a long time most researchers neither realized (nor believed) (1) that OA was fully within their reach, nor (2) that it could bring them substantial benefits.

(1) **Reachability of OA.** That OA was fully within researchers' reach had in fact already been demonstrated decades earlier, first by the computer scientists who invented Unix and the Internet, and immediately began storing and sharing their papers on 'anonymous FTP sites' (in the '80's, and then on websites, once the web was invented). Next, some branches of physics -- notably high energy physics (in which there had already been a culture, even in the paper era, of systematically sharing one another's prepublication preprints) -- made the natural transition to first sharing preprints via email and then via the web : In the early '90s, physicists began self-archiving their papers in electronic form ('eprints') -- both before (preprints) and after peer-review (postprints) -- in a central web archive (long called XXX and eventually [Arxiv]()). In computer science, meanwhile, where self-archiving had been going on even longer, but on authors' local FTP and websites rather than centrally, a central harvester of those local websites, [Citeseer](), was created in the late '90s that not only gathered together all the computer science papers it could trawl from the web, but extracted and linked their reference lists, generating a rudimentary citation count for each article (Giles et al. 1999). Shortly thereafter, [Citebase]() was created to do the same (and more) for the contents of Arxiv (Brody 2003). (We will return to Citebase, shortly.)

(2) **Benefits of OA.** But apart from these two bursts of spontaneous self-archiving in computer science and some areas of physics, other disciplines did not pick up on the power and potential of the online medium for enhancing the usage of their work. Lawrence (one of the co-inventors of Citeseer) published a study in Nature in 2001, showing that articles that were made freely available on the Web were cited more than twice as much as those that were not ; yet most researchers still did not rush to self-archive. The finding of an OA citation impact advantage was soon extended beyond computer science, first to physics (Harnad & Brody 2004), and then also to all 10 of the biological, social science, and humanities disciplines so far tested (Hajjem et al 2005) ; yet the worldwide spontaneous self-archiving rate continued to hover around 15%.

If researchers themselves were not very heedful of the benefits of OA, however, their institutions and research funders – co-beneficiaries of their research impact – were: To my knowledge, the department of Electronics and Computer Science (ECS) at University of Southampton was the first to mandate self-archiving for all departmental research articles published: These had to be deposited in the department's own [Institutional Repository (IR)]() (upgraded using the first free, open source software for creating OA IRs, likewise created at Southampton and now widely used worldwide).

Southampton was persuaded to do what it did in part because of something unique to the United Kingdom : The [Research Assessment Exercise (RAE)](), in which research performance indicators from every department in every UK university are evaluated and ranked every six years or so, with the departments receiving substantial top-sliced research funding as a function of their RAE rank.

**The UK Research Assessment Exercise (RAE).** The RAE is a very cumbersome, time-consuming and expensive undertaking, for the researchers as well as the assessors. It requires preparing and submitting piels and piles of paper containing indicators of research productivity and performance (funding, students, publications, applications, industrial spin-offs, etc.) along with each researcher's four best papers, which are then 'peer-reviewed' by an RAE panel for each discipline. Of course, these papers are already published, hence have already been peer-reviewed; moreover, those that were published in the better journals had already been peer-reviewed by the best experts in the world in their field, not a generic RAE panel. But the interesting thing about the RAE outcome was that although (for no particular reason) it explicitly forbade citation counts among its performance indicators, the RAE rankings nevertheless turned out to be highly correlated with the total citation counts for the submitted researchers, and this was found to be true in every one of the RAE disciplines for which the correlation was tested (Oppenheim 1996).

The question then arises : If the RAE ranks that result from this complicated and time-consuming process are highly correlated with citations that are not even explicitly counted, why not just count the citations? If the correlation is high enough, the cumbersome process can be dropped, or at least simplified. Professor [Charles Oppenheim]() of Loughborough University (who did most of the RAE/citation correlation studies), the University of Southampton research group that did most of the [OA citation advantage]() studies, and many others in the UK, accordingly urged that the RAE should be dropped or simplified in favor of metrics, and it was subsequently decided to do so: The next RAE in 2008 will be a parallel exercise, conducted the old, complicated way, alongside a new, stream-lined, metrics-based way (Harnad 2006). But another complication has arisen:

In many of the hard-science disciplines there was a metric that correlated with the RAE outcome even more highly than citations: *prior research funding*. One could hence make an even stronger argument for basing the RAE rank entirely on prior research funding in those disciplines (where the correlation was close to 1.0). The problem, however, is that research funding <u>is</u> explicitly submitted as a performance indicator. So its tight correlation with the RAE outcome could well be because the RAE panels, in making their evaluations, were strongly influenced (indeed biassed) by the amount of prior funding received – perhaps more influenced than by other factors, such as, for example, the quality of the submitted papers (which the panels were not necessarily even best-qualified to judge). Citations, in contrast, were not submitted or counted; so their correlation with the outcome, though not as high as that of prior funding, was 'unbiassed'. (Moreover, the RAE is meant to be an independent, top-sliced component of UK research funding, alongside classical competitive proposal-based funding, in a *dual* funding system. If prior funding were given too heavy a weight in the RAE ranking, that would merely amount to collapsing the dual funding system into a single proposal-based system, and just adding a multiplier effect to the proposal-based funding, amplifying any Matthew Effect, with the rich just getting richer and the poor poorer.)

There is a lesson to be learned here, and we will return to it, but first, back to the problem of the low spontaneous OA self-archiving rate worldwide, despite the evidence that it doubles citations. Part of the reasoning that drove Southampton's ECS department to adopt the world's first self-archiving mandate was that it would increase the visibility and usage of the department's research, thereby increasing its citation impact, and, in turn, the department's RAE rank. But there was also a lot of enthusiasm in the department for OA self-archiving, arising from studies based on citation-linking the articles deposited in the Physics Arxiv, through [Citebase](), the scientometric search engine created by Tim Brody, then a doctoral student in the department (Brody 2003; Harnad et al. 2003; Harnad & Brody 2004).

**Citebase.** In conjunction with citation data from the ISI database (from a leased ISI CD-ROM), Citebase was used to compare the citation counts for articles in the same physics journals and years that were and were not self-archived in Arxiv. Lawrence's OA citation advantage in computer science was confirmed in physics. Brody then went on to compare the download counts for articles in Arxiv within the first six months after they were deposited, and their citation counts a year or more later – and found a significant correlation : Early citation counts predict later citation counts, but download counts predict citations even earlier. A download/citation correlator was designed that allowed the user to set the time-window for calculating the correlation between downloads and citations (Brody et al 2006).

Citebase accordingly began adding more and more metrics on which it could rank the results of searches, including (for either the paper or the author): dates (age), citations, downloads, hub scores, authority scores, co-citedness scores. These rankings were of course for demonstration purposes only, and unvalidated. They were also very noisy, as authors' names were not uniquely sorted, the download data only came from the UK mirror site of Arxiv (as the primary US Arxiv site did not allow us to analyze their download statistics), not all articles in Arxiv had been successfully linked, and of course Arxiv's coverage of physics is not complete. Nevertheless, Citebase demonstrated the potential power of citation-based navigation as well as (univariate) scientometric ranking.

**A Scientometric Ranking Engine for the RAE.** The natural next step – apart from increasing Citebase's coverage – is to add more of the 'vertical' metrics that currently allow univariate ranking, one variable at a time, but to turn them instead into 'horizontal' multivariate metrics, in a dynamic multiple-regression engine, in which each of the individual vertical predictors can be given a weight and combined in a multivariate linear equation. This can of course be done by simply picking one (or more) of the univariate metrics themselves, and using it as the criterion. That allows some rudimentary assignment of the beta weights on each remaining predictor. Or the weights can be hand-calibrable, so the user can explore the ranking yielded by different (normalized) hand-adjusted weights.

But the best solution of all is to provide an external criterion for validating the weights on the battery of predictor metrics. And there is, in the case of RAE 2008, a natural criterion, namely, the ranks generated by the parallel panel-based ('peer review') exercise itself. For all submitted authors (or papers), a Citebase-like multiple regression engine can be fed their publication counts, their articles' journal impact factors, citation counts, co-citation counts, citation chronometrics (age, growth, latency to peak, decay rate), hub/authority scores, h-index, prior funding, student counts, co-authorship scores, endogamy/exogamy scores, textual proximity scores, download and co-download counts and chronometrics, and more. The multiple regression of all those predictor metrics onto the panel rankings as the validation criterion will generate the default beta weights (different in each discipline) for each metric (Harnad 2006). This will give an indication of the validity of the predictor metrics as a whole, relative to the panel-based rankings as the criterion. It will also allow adjustments, to calibrate the weights on the predictors so as to fine-tune the outcome or eliminate biases (e.g., to reduce or remove the Matthew Effect of prior funding; or there may be other factors to amplify, reduce, minimize, or eliminate). The battery could also be factor-analyzed to look for a smaller number of underlying factors – possibly even a 'G-factor', if there is one.

The UK RAE is merely one glimpse of the possibilities opened up by OA scientometrics (Shadbolt et al. 2006). There is no more reason for these metrics to remain inaccessible to all users in the online age than for the research articles themselves to remain inaccessible. The data-mining potential of an OA corpus is enormous, not just for research evaluation by performance assessors, but for search and navigation by reseacher-users, students, and even the general public. Historians of knowledge, as well as analysts trying to predict the future course of knowledge will gain a wealth of powerful new tools to do so. The only thing missing is the primary OA database: the articles themselves. The main allies in demonstrating to the research community the benefits of providing that primary OA database today are the self-same scientometric tools that are waiting to mine it.